\documentstyle[pra,aps,multicol]{revtex}
\renewcommand{\narrowtext}{\begin{multicols}{2} \global\columnwidth20.5pc}
\renewcommand{\widetext}{\end{multicols} \global\columnwidth42.5pc}
\begin{document}
\draft
\title{Quantum Clone and States Estimation for $n$-state System}
\author{Chuan-Wei Zhang, Chuan-Feng Li*, and Guang-Can Guo$^{\sharp }$}
\address{Laboratory of Quantum Communication and Quantum\\
Computation and Department of Physics,\\
University of Science and Technology of China,\\
Hefei 230026, People's Republic of China\vspace*{0.3in}}
\maketitle

\begin{abstract}
\baselineskip12ptWe derive a lower bound for the optimal fidelity for
deterministic cloning a set of $n$ pure states. In connection with states
estimation, we obtain a lower bound about average maximum correct states
estimation probability.

\end{abstract}

\vskip 1.0cm

\narrowtext
\baselineskip12pt

Quantum no-cloning theorem \cite{Woot82} has prohibited cloning and
estimating an arbitrary quantum state exactly by any physical means in a
consequence of linearity of quantum theory. The unitarity of quantum theory
does not allow to clone (identify) no-orthogonal states though orthogonal
states can be cloned (identified) perfectly \cite{Yuen86}. However, clone
and estimation of quantum states with a limited degree of success are always
possible. Universal quantum cloning machine (UQCM) \cite
{Buz96,Gis97,Wer98,Buz98,Cerf00} acts on any unknown quantum state and
produce optimal approximate copies. This machine is called universal because
it produces copies that are state-independent. State-dependent quantum
cloning machines is designed to clone states belonging to a finite set and
may be divided into two main categories: deterministic \cite{Hil97},
probabilistic \cite{Duan98,Zhang00,Pati99} and hybrid \cite{Che99}.
Deterministic state-dependent cloning machine generates approximate clones
with probability 1. Deterministic exact clone violates the no-cloning
theorem, thus perfectly clone must be probabilistic. Probabilistic quantum
cloning machines can clone states perfectly, though the success probability
cannot be unit all the time. It is shown that a set of non-orthogonal states
can be probabilistically cloned if and only if the states are linearly
independent. Hybrid clone interpolates between deterministic and
probabilistic ones, that is, the copies (not exact) are better than those in
deterministic clone, but the success probability (less than 1) is greater
than probabilistic exact clone. Universal quantum states estimation were
considered in Ref. \cite{Mas95}, given $M$ independent realizations. What's
more, we \cite{Zhang99} have discussed general states discrimination
strategies for state-dependent system.

Optimal results for two-state deterministic clone have been obtained in Ref. 
\cite{Hil97,Che99}. In this letter we consider deterministic clone for a set
of $n$ pure states $\left\{ \left| \psi _i\right\rangle ,i=1,2,...,n\right\} 
$. When $\left| \psi _i\right\rangle $ are non-orthogonal, they cannot be
cloned perfectly. What we require is that the final states should be most
similar as the target states, that is, the fidelity between the final and
target states should be optimal. We derive a lower bound for the optimal
fidelity of the cloning machine. Applying it to states estimation, we obtain
the lower bound about average maximum correct identification probability in
deterministic states estimation.

A quantum state-dependent cloning device is a quantum machine which performs
a prescribed unitary transformation on an extended input which contains $M$
original states in system $A$ and $N-M$ blank states in system $B$ with $N$
output copies. The unitary evolution transfers states as follows 
\begin{equation}
U\left| \psi _i^M\right\rangle _A\left| \Sigma ^{N-M}\right\rangle _B=\left|
\alpha _i\right\rangle \text{,}  \label{1}
\end{equation}
where $\left| \psi _i^M\right\rangle _A=\left| \psi _i\right\rangle
_1\otimes ...\otimes \left| \psi _i\right\rangle _M$ are the $M$ original
states, $\left| \Sigma ^{N-M}\right\rangle _B$ are the blank states and $%
\left| \alpha _i\right\rangle $ are the output cloned states. The $n\times n$
inter-inner-products of Eq. (1) yield the matrix equation\footnote{%
We notice the preserving inner-product property of unitary transformation,
that is, if two sets of states $\left\{ \left| \phi _1\right\rangle ,\left|
\phi _2\right\rangle ,...,\left| \phi _n\right\rangle \right\} $ and $%
\left\{ \left| \widetilde{\phi }_1\right\rangle ,\left| \widetilde{\phi }%
_2\right\rangle ,...,\left| \widetilde{\phi }_n\right\rangle \right\} $
satisfy the condition $\langle \phi _i\mid \phi _j\rangle =\langle 
\widetilde{\phi }_i\mid \widetilde{\phi }_j\rangle $, there exists a unitary
operate $U$ to make $U\left| \phi _i\right\rangle =\left| \widetilde{\phi }%
_i\right\rangle $ ($i=1,2,...,n$).} 
\begin{equation}
X^{(M)}=\tilde{X}\text{,}  \label{2}
\end{equation}
where $n\times n$ matrices $\tilde{X}=\left[ \left\langle \alpha _i|\alpha
_j\right\rangle \right] $, $X^{(M)}=\left[ \left\langle \psi _i|\psi
_j\right\rangle ^M\right] $.

We require a figure of merit to characterize how closely our copies $\left|
\alpha _i\right\rangle $ resemble exact copies $\left| \psi
_i^N\right\rangle $. Denoting the priori probability of the state $\left|
\psi _i^M\right\rangle $ by $\eta _i$, one interesting measure of the final
states is the global fidelity introduced by Bru$\beta $ {\it et al.} \cite
{Hil97}, which is defined formally as 
\begin{equation}
F=\sum_{i=1}^n\eta _i\left| \langle \alpha _i\left| \psi _i^N\right\rangle
\right| ^2.  \label{3}
\end{equation}
As a criterion for optimality of the state-dependent cloner, the unitary
evolution $U$ should maximize the global fidelity $F$ of $n$ final states $%
\left| \alpha _i\right\rangle $ with respect to the perfect cloned states $%
\left| \psi _i^N\right\rangle $. We focus here on the global fidelity since
it has an important interpretation in connection with states estimation \cite
{Che99}.

Now the remained problem is to find the maximum value of the fidelity $F$,
which means optimal clone. It is equivalent to the problem of maximizing $F$
under the condition of Eq. (2). This problem is a nonlinear programming and
fairly difficult to solve. Nevertheless a lower bound of the optimal
fidelity could still be derived by adopting an auxiliary function $%
F^{^{\prime }}$, which is defined as 
\begin{equation}
F^{^{\prime }}=\sum_{i=1}^n\eta _i\left| \left\langle \psi _i^N\mid \alpha
_i\right\rangle \right| .  \label{4}
\end{equation}
Such function also describes how closely our output copies resemble exact
copies. There exists a bound between $F$ and $F^{^{\prime }}$ (see below,
inequality (9)), therefore a bound for $F$ may be obtained by optimizing $%
F^{^{\prime }}$.

We find that the optimal output states $\left| \alpha _i\right\rangle $ must
lie in the subspace spanned by the exact clones $\left| \psi
_i^N\right\rangle $. This conclusion may be easily come to with the method
of Lagrange Multipliers (please refer to \cite{Hil97}, where $n=2$) and here
we omit the proof.

If a set of states $\left| \tilde \alpha _i\right\rangle $ fulfil Eq. (2),
that is, $X^{\left( M\right) }=\tilde X=\left[ \left\langle \tilde \alpha _i|%
\tilde \alpha _j\right\rangle \right] $, there must exist a unitary
transformation $V$ satisfies $V\left| \tilde \alpha _i\right\rangle =\left|
\alpha _i\right\rangle $, thus we can vary $V$ to optimize $F^{^{\prime }}$
with chosen states $\left| \tilde \alpha _i\right\rangle $. Suppose $\left|
\left\langle \psi _i^N\mid \alpha _i\right\rangle \right| =\lambda
_i\left\langle \psi _i^N\mid \alpha _i\right\rangle $ with $\lambda _i\in
\left\{ \pm 1\right\} $ in the optimal situation (the determination of $%
\lambda _i$ will be shown in later part), the optimal $F^{^{\prime }}$ is 
\begin{equation}
F_{opt}^{^{\prime }}=\max_VF^{^{\prime }}=\max_V\left| \sum_{i=1}^n\eta
_i\lambda _i\left\langle \psi _i^N\right| V\left| \tilde \alpha
_i\right\rangle \right| .  \label{5}
\end{equation}

Choose $n$ orthogonal states $\left| \chi _i\right\rangle $ which span a
space ${\cal H}$ and the space spanned by $\left| \psi _i^N\right\rangle $
is a subspace of ${\cal H}$\footnote{%
We consider space $\left\{ \left| \psi _i^N\right\rangle
,i=1,2,...,n\right\} $ may be a subspace of ${\cal H}$ since $\left| \psi
_i\right\rangle $ may be linear-dependent and cannot span a $n$-dimensional
Hilbert space.}. Set $\left| \tilde{\alpha}_i\right\rangle
=\sum\limits_{j=1}^na_{ij}\left| \chi _j\right\rangle $, $\left| \psi
_i^N\right\rangle =\sum\limits_{j=1}^nb_{ij}\left| \chi _j\right\rangle $ on
the orthogonal bases $\left| \chi _i\right\rangle $, $i=1,2,...,n$, we get 
\begin{equation}
F_{opt}^{^{\prime }}=\max_V\left| tr\left( \eta \lambda BVA^{+}\right)
\right| =\max_V\left| tr(VO)\right| =tr\sqrt{O^{+}O},  \label{6}
\end{equation}
where $A=\left[ a_{ij}\right] $, $B=\left[ b_{ij}\right] $, $\eta =diag(\eta
_1,\eta _2,...,\eta _n)$, $\lambda =diag(\lambda _1,\lambda _2,...,\lambda
_n)$, $O=A^{+}\eta \lambda B$. We have used the freedom in $V$ to make the
inequality as tight as possible. To do this we have recalled \cite{Joz94}
that $\max\limits_V\left| tr(VO)\right| =tr\sqrt{O^{+}O}$, where $O$ is any
operator and the maximum is achieved only by those $V$ such that 
\begin{equation}
VO=e^{i\nu }\sqrt{O^{+}O},  \label{7}
\end{equation}
where $\nu $ is arbitrary. Generally, we choose $\nu =0$.

As we require above, $\lambda _i$ should satisfy $\lambda _i\left\langle
\psi _i^N\right| V\left| \tilde \alpha _i\right\rangle \geq 0$. This
condition can be represented as $\left\langle \chi _i\right| \lambda
BVA^{+}\left| \chi _i\right\rangle \geq 0$, which means the diagonal
elements of matrix $\lambda BVA^{+}$ should be positive. Since $\lambda
_i\in \left\{ \pm 1\right\} $, a simple method to determine $\lambda _i$ is
to enumerate the $2^n$ possible results of $\lambda =diag(\lambda _1,\lambda
_2,...,\lambda _n)$ and verify which one fulfils above inequality. With a
chosen basis $\left| \chi _i\right\rangle $, matrix $A$, $B$ can be given by
equations $A^{+}A=X^{\left( M\right) }$ and $B^{+}B=X^{\left( N\right) }$
respectively, $V$ can be represented with parameters $\lambda _i$, thus
above postcalculation method can determine matrix $\lambda $ and then give
the maximum $F_{opt}^{^{\prime }}$. According to Eq. (6), we obtain a tight
upper bound for the function $F^{^{\prime }}$, 
\begin{equation}
F^{^{\prime }}\leq tr\sqrt{B^{+}\lambda \eta X^{\left( M\right) }\eta
\lambda B}.  \label{8}
\end{equation}

The fidelity $F$ of the cloning machine is constrained by the following
inequality
\begin{eqnarray}
F &=&\left( \sum_{i=1}^n\eta _i\left| \langle \alpha _i\left| \psi
_i^N\right\rangle \right| ^2\right) \left( \sum_{i=1}^n\eta _i\right) 
\label{9} \\
&\geq &\left( \sum_{i=1}^n\eta _i\left| \langle \alpha _i\left| \psi
_i^N\right\rangle \right| \right) ^2=\left( F^{^{\prime }}\right) ^2, 
\nonumber
\end{eqnarray}
where the equation is met if and only if $\left| \langle \alpha _i\left|
\psi _i^N\right\rangle \right| $ are constant. Obviously $F$ is not always
optimal even if $F^{^{\prime }}$ is optimal. However optimal $F$ should be
greater than or equal to $\left( F_{opt}^{^{\prime }}\right) ^2$. When $n=2$
and $\eta _1=\eta _2$, equation in Ineq. (9) is satisfied and gives the
optimal results, which has been provided in Ref. [8,12].

State-dependent clone has a close connection with states estimation in the
limit as $N\rightarrow \infty $. Given infinite copies of $n$ non-orthogonal
states, we can discriminate them exactly with probability 1. On the other
hand, if we can discriminate $n$ states, we can obtain infinite copies.
There are two ways in which an attempt to discriminate between
non-orthogonal states; it can give either an erroneous or an inconclusive
result \cite{Zhang99}. In the following we will consider a strategy without
inconclusive results using above results in the limit as $N\rightarrow
\infty $. In fact, since the optimal output states $\left| \alpha
_i\right\rangle $ lie in the subspace spanned by the exact clones $\left|
\psi _i^N\right\rangle $, Eq. (1) may be rewritten as 
\begin{equation}
U\left| \psi _i^M\right\rangle \left| \Sigma ^{N-M}\right\rangle
=\sum_{j=1}^nc_{ij}\left| \psi _j^N\right\rangle \text{,}  \label{10}
\end{equation}
where $c_{ij}=\left\langle \psi _j^N\mid \alpha _i\right\rangle $. If $%
N\rightarrow \infty $, $\left\{ \left| \psi _j^N\right\rangle
,j=1,2,...,n\right\} $ are orthogonal. After the evolution, the cloning
system is measured and if $\left| \psi _j^\infty \right\rangle $ is
obtained, the original state is estimated as $\left| \psi _j^M\right\rangle $%
. The states estimation is correct with probability $\left| c_{ii}\right| ^2$
when $j=i$. If $j\neq i$, errors occur with probability $\sum\limits_{j\neq
i}\left| c_{ij}\right| ^2$. The inter-inner products of Eq. (10) give the
matrix equation in the limit $N\rightarrow \infty $, 
\begin{equation}
X^{(M)}-EE^{+}=0,  \label{11}
\end{equation}
where $E=\left[ c_{ij}\right] $. The diagonal elements is corresponding to
the probabilities of correct states estimation while non-diagonal elements
to those of error. This equation describes the bound between the maximum
probabilities of correct discrimination and those of incorrect one. In fact,
this result is a special case of that we have derived in \cite{Zhang99}. In
Ref. \cite{Zhang99}, we have consider two possible ways in which an attempt
to discriminate between non-orthogonal states can fail, by giving either an
erroneous or an inconclusive result. Above strategy just gives an erroneous
result with some probability. Our principal result in Ref. \cite{Zhang99} is
the matrix inequality which prescribes the bound among the probabilities of
correct, error and inconclusive discrimination results. Such bound may have
intriguing implications for quantum communication theory and cryptography 
\cite{Ben97} since it offers a potential eavesdropper increased flexibility
by a compromise between inconclusive and erroneous results.

An important optimality criterion of the states estimation is the average
maximum correct probability, that is, $P=\sum_i\eta _i\left| c_{ii}\right|
^2=F$ in the limit $N\rightarrow \infty $\footnote{%
It is the reason why we choose the definition of $F$ as that in Eq. (3).}.
In this situation $\left| \psi _j^N\right\rangle $ are orthogonal, thus
matrix $B=I_n$. Applying Eq. (8) and (9), we obtain 
\begin{equation}
P=\sum_i\eta _i\left| c_{ii}\right| ^2\geq \left( tr\sqrt{\lambda \eta
X^{(M)}\eta \lambda }\right) ^2.  \label{12}
\end{equation}
Such $F$ is not always optimal bound of the average maximum probability of
correct states estimation, however, the optimal one is always greater than $%
\left( tr\sqrt{\lambda \eta X^{(M)}\eta \lambda }\right) ^2$.

We note that above bound about $F$ and $P$ have the meaning in average. They
describe the optimality approach to the final states we can reach in average
of the $n$ initial states and does not mean the best for each initial state.
However, since we do not know which one the initial state is in the clone or
estimation process, such average may be the most important value to describe
the efficiencies of cloning (estimating) machines.

In summary, we have derived a lower bound for the optimal fidelity for the
state-dependent quantum clone. In connection with states estimation, we
obtained the matrix inequality which describes the bound between the maximum
probabilities of correct discrimination and those of incorrect one. A lower
bound about average maximum probability of correct identification has also
been presented. Our results give some bounds which the optimal cloner and
states estimation can be better than in average, however, we have not found
a limit which optimal cloner can reach at most. It is still an open question
needed to be explored.

{\bf ACKNOWLEDGMENTS}: This work was supported by the National Natural
Science Foundation of China.

* Electronic address: cfli@ustc.edu.cn

$^{\sharp }$ Electronic address: gcguo@ustc.edu.cn

%TCIMACRO{\TeXButton{widetext}{\widetext}}
%BeginExpansion
\widetext%
%EndExpansion

\end{document}